\documentclass{emulateapj}
\usepackage{apjfonts}

\begin{document}

\title{Discovery of a spatially extended GeV source in the vicinity of the TeV halo candidate 2HWC J1912+099: a TeV halo or supernova remnant ?}
\author{Hai-Ming Zhang\altaffilmark{1,4}, Shao-Qiang Xi\altaffilmark{1,4}, Ruo-Yu Liu\altaffilmark{1,2}, Yu-Liang Xin\altaffilmark{3}, Siming Liu\altaffilmark{3}, Xiang-Yu Wang\altaffilmark{1,4}}
\altaffiltext{1}{School of Astronomy and Space Science, Nanjing University, Nanjing 210023, China; xywang@nju.edu.cn; ryliu@nju.edu.cn}
\altaffiltext{2}{Deutsches Elektronen Synchrotron (DESY), Platanenallee 6, D-15738 Zeuthen, Germany;}
\altaffiltext{3}{Purple Mountain Observatory and Key Laboratory of Radio Astronomy, Chinese Academy of Sciences, Nanjing 210008, China}
\altaffiltext{4}{Key laboratory of Modern Astronomy and Astrophysics (Nanjing University), Ministry of Education, Nanjing 210023, China}

\begin{abstract}
Observations  by HAWC and Milagro have detected spatially extended TeV sources surrounding middle-aged ($t\sim100-400 \,{\rm kyr}$) pulsars like Geminga and PSR B0656+14, which have been named  ``TeV Halos'', representing very extended TeV pulsar wind nebulae (PWNe) powered by relatively old pulsars. A few more HAWC-detected sources have been suggested to be TeV halo candidates. In this paper, we search for possible GeV counterparts of three TeV halo candidates with Fermi Large Area Telescopes. We detect a new spatially extended GeV source in the vicinity of the TeV halo candidate 2HWC J1912+099, which is also detected by HESS (HESS J1912+101). We find that the size of the GeV source is significantly larger than that of the TeV emission measured by HESS, and a spatial template characteristic of a PWN can fit the GeV data. We suggest that the GeV source is an extended PWN powered by the central middle-aged pulsar PSR J1913+1011. This discovery favors the TeV halo scenario for  the TeV source 2HWC J1912+099 (HESS J1912+101), although the possible shell-like morphology measured by HESS  challenges this interpretation. Alternatively, the TeV emission could be dominated by a supernova remnant (SNR) via the hadronic process. Future multi-wavelength observations of the source and more precise measurements of the spatial profile of the TeV emission will be useful to distinguish between the two scenarios.

\end{abstract}


\section{Introduction}           
\label{sect:intro}
Observations from Milagro (Abdo et al. 2009), along with recent observations by HAWC (Abeysekara et al. 2017a), have revealed extended TeV emissions (i.e., TeV halos) surrounding Geminga and PSR B0656+14. The angular sizes of the TeV halos are much larger than the X-ray pulsar wind nebulae (PWNe). The measurements of surface brightness profile of these TeV halos suggest inefficient diffusion of particles from the sources, giving rise to a debate on the pulsar interpretation of the cosmic-ray positron excess (Abeysekara et al. 2017b). Xi et al. (2019) argued that GeV observations provide more direct constraints on the positron density in the TeV nebulae in the energy range of 10-500 GeV and hence on the origin of the observed positron excess. Motivated by this, they searched for GeV emission from the two TeV halos with the Fermi Large Area Telescope (LAT). No convincing GeV counterparts are detected from these two TeV halos which suggests a relatively low density of GeV electrons/positrons in the TeV halos, thereby constraining their contribution to the positron excess (Xi et al. 2009). {Considering the proper motion of the Geminga pulsar and one-zone diffusion spatial templates, which leads to a very large size for the source,  Di Mauro et al. (2019) claimed the detection of GeV emission from the Geminga TeV halo. However, the large size could lead to serious contamination from the background, so the result is still uncertain. }

A few more TeV halo candidates have been suggested by Linden et al. (2017), based on the HAWC source catalog (Abeysekara et al. 2017a). These include 2HWC J1912+099,  2HWC J1831-098 and 2HWC J2031+415 (Abeysekara et al. 2017a). Motivated by these observations, we attempt to search for possible GeV counterparts of these TeV halo candidates.  2HWC J1912+099 is spatially coincident with HESS 1912+101, which is found in the Galactic plan survey by HESS. HESS 1912+101 is suggested to be a possible supernova remnant based on the shell-like morphology of the TeV emission (H.E.S.S collaboration 2018). However, the suggested size of the TeV emission measured by HAWC is different from the HESS source (Abeysekara et al. 2017a),  which may be  due to different field-of-view and sensitivities of the two telescopes.

The area of HESS J1912+101 was covered by the NRAO/VLA Sky Survey at 1.4 GHz (Condon et al. 1998) and the new Multi-Array Galactic Plane Imaging Survey (MAGPIS; Helfand et al. 2006), but no obvious counterpart to the TeV source was found (H.E.S.S collaboration 2018). Recently, Reich \& Sun (2019) analysed the Sino-German Urumqi $\lambda6$ cm survey and find a partial shell of possible excessive polarisation at $\lambda6$ cm at the periphery of HESS J1912+101, but they could not separate the shell's total intensity signal from the confusing intense diffuse emission from the inner Galactic plane. To be conservative, since the significance of polarization is low,  we use their total flux density of $2.5\pm 1.0{\rm Jy}$, which is inferred from the polarization flux, as an upper limit at $\lambda6$ cm (Reich \& Sun 2019).  Archival Chandra data from observations targeting at PSR J1913+1011 (and only covering the central region of HESS J1912+101) were analyzed by Chang et al. (2008) to  search for X-ray counterparts, but no convincing X-ray counterpart was detected. As no low-energy (radio and X-rays) counterparts are found so far,  the nature of this source is quite uncertain.

In \S 2, we perform an analysis on the \textit{Fermi-LAT} data towards the region of 2HWC J1912+099 (HESS 1912+101).  We focus on the spatial extension analysis of the GeV emission using different spatial templates. In \S 3, we consider a physically-motivated template, i.e., the GeV emission is produced by diffusing electrons  as in the TeV halos (hereafter named "diffusion" templates). In \S 4, we present the spatial analysis of the other two TeV halo candidates, 2HWC J1831-098 and 2HWC J2031+415. Finally , we give  discussions in \S 5.

\section{Fermi/LAT Data Analysis}
The LAT on board the \textit{Fermi Gamma-Ray Space Telescope} has continuously monitored the sky since 2008 and scans the entire sky every 3 hours (Atwood et al. 2009). For this work we use the \texttt{Pass 8} SOURCE data taken from 2008 August 4 to 2019 January 19 to study the extended gamma-ray emission around HESS J1912+101. We select the $\gamma$-ray events in the $10-500 \, \rm GeV$ energy range, using standard data quality selection criteria "$(DATA\_QUAL > 0)  \&\& (LAT\_CONFIG == 1)$".
Since we are only considering high energy ($>$10 GeV) $\gamma$-ray events, for which the point-spread function(PSF) is relatively good, the maximum zenith angle is set to be 105$\degr$.
Data within a $14\degr \times14\degr$ radius of interest (ROI) centered on the position of HESS J1912+101  are binned in 12 logarithmically spaced bins in energy and  a spatial binning of 0.1$\degr$ per pixel is used. We utilize the the publicly available software \textit{fermitools} (ver. 1.0.0). The \textit{$P8R3\_SOURCE\_V2$} set of instrument response functions (IRFs) is used.

For the background model, we include the diffuse Galactic interstellar emission (IEM, $gll\_iem\_v07.fits$) and isotropic emission ("$iso\_P8R3\_SOURCE\_V2\_v1.txt$" ) templates released by Fermi Science Support Center (FSSC)\footnote{{http://fermi.gsfc.nasa.gov/ssc/data/access/lat/BackgroundModels.html}}, as well as individual gamma-ray sources listed in  the Fourth Catalog of \textit{Fermi}-LAT Sources (4FGL; The Fermi-LAT collaboration 2019). Considering that the background point source 4FGL J1913.3+1019 is reported to be associated with a radio pulsar,  PSR J1913+1011, which is only $\sim0.140\degr$ away (RA=288.335$\degr$,DEC=10.19$\degr$; Morris et al. 2002), we re-locate the position of 4FGL J1913.3+1019 to that of PSR J1913+1011 in the analysis.

In subsequently analysis, the normalization and spectral parameters of the discrete gamma-ray sources within 5$\degr$ in the background model are left free. We also free the normalizations of the isotropic and Galactic components. Note that the typical cutoff energies of the sources shaped by the LogParabola function are  smaller than 10 GeV,  we thus free only the normalization of the LogParabola function in order to get a convergence of fitting.

We search for possible GeV emission around the region of HESS J1912+101. First, we perform a background-only fitting and obtain the test statistic (TS) map, as shown in Fig. 1. The TS value is defined as TS=2($\ln\mathcal{L}_{1}-\ln\mathcal{L}_{0})$, where $\mathcal{L}_{0}$ is the likelihood of background (null hypothesis) and $\mathcal{L}_{1}$ is the likelihood of the hypothesis for adding a source. An obvious excess is located in the region of HESS J1912+101. Then, we create a series of spatial templates to test the extension of the excess emission. To define a source to be extended, $TS_{ext}\geq16$ is required ($TS_{ext}=16$ corresponds to a formal 4$\sigma$ significance; Ackermann et al.2017), where  $TS_{ext}=2(\ln\mathcal{L}_{ext}-\ln\mathcal{L}_{ps})$, with $\mathcal{L}_{ext}$ and $\mathcal{L}_{ps}$ being the likelihoods of hypothesis for adding an extended source and a point-like source, respectively (Lande et al. 2012).

\subsection{Single point-like source model}
First, we consider a point-like source model. We add a point-like source  at the position of the peak test statistic (TS) value into our background model, and optimize the localization using the \textit{gtfindsrc} tool.
The derived best-fit location of the excess above 10 GeV is (288.405$\degr$, 10.361$\degr$)$\pm0.028\degr$. The significance of the excess $\gamma$-ray emission is TS = 45.3 (6.7$\sigma$). The spectrum and flux in this model are shown in Table 1.

\subsection{HESS map template}

To check if the spatial distribution of the excess $\gamma$-ray emission traces that of the observed TeV emission measured by HESS (HESS J1912+101), we consider a template that reflects the TeV emission morphology measured by HESS, i.e.,  a three-dimensional spherical shell, homogeneously emitting between $R_{in}$ and $R_{out}$ and projected onto the sky, with  $R_{out}=0.49\degr$ and  $R_{in}=0.32\degr$ (H.E.S.S collaboration 2018; see also the red "dashed circles" in Fig. 1). The significance of the excess $\gamma$-ray emission is TS = 43.2 (6.6$\sigma$). This TS value indicates that the HESS map model does not improve over the point source model. This seems to indicate that there  is a peak in the GeV emission near the center of the profile, in contrast to a ring-like distribution.

\subsection{Uniform disk template}
To investigate whether the $\gamma$-ray excess is an extended source spatially, we first consider an uniform disk template. We select the best-fit position of the point-like source as the center of the disk, and vary the disk radii from 0.1$\degr$ to 2.0$\degr$ in steps of 0.1$\degr$ to search for the best-fit radius. We also vary the radii in steps of 0.05$\degr$ around the best-fit radius found in the first step to get a more precise radius. The best-fit  is found at a disk radius of $R_{disk}=0.85\degr\pm0.08\degr$ with TS=68.5 (8.3$\sigma$), as shown in Fig. 2.  Since we have $TS_{ext}=23.1$ relative to a single point source model, the GeV source is spatially extended.  The radius of GeV emission is larger than that of the TeV emissions measured by HAWC and HESS, which are about $0.7\degr$ and $0.5\degr$, respectively (Abeysekara et al. 2017a; H.E.S.S collaboration 2018).

\subsection{2D Gaussian template}

The other model for studying the source extension in standard \textit{Fermi-LAT} analysis is the two-dimension (2D) Gaussian model.
We set the center of 2D Gaussian  at the best-fit position of the point source model, and vary  $\sigma$ from 0.1$\degr$ to 1.3$\degr$ in steps of 0.1$\degr$ to search the best-fit $\sigma$. Also we consider a step of 0.01$\degr$ around the best-fit $\sigma$ found in the first step. We define the source size as the radius containing 68$\%$ of the intensity, $R_{68}=1.51\sigma$, as suggested by Lande et al. (2012).
The best-fit is found at $\sigma=0.42\degr\pm0.03\degr$, corresponding to 68\% containment radius $R_{68}=0.63\pm0.05\degr$, with TS=76.7 (8.8$\sigma$), as shown in Fig. 2. The significance of extension in the 2D Gaussian model is $TS_{ext}=31.4$ ($5.6\sigma$) relative to a single point source model.

\section{Diffusion templates in the TeV halo scenario}
 The fact that the size of the GeV source is significantly larger than that of the TeV source is consistent with the scenario of the counterpart of a TeV halo (i.e., an extended GeV PWN), considering that GeV-emitting electrons have longer cooling time so they diffuse to a larger extent. Although the shell-like morphology measured by HESS is not expected in the TeV halo scenario, we note, however, that the morphology and size measured by HAWC are different, so it is still unclear whether the TeV halo scenario can be excluded. On the other hand, interpreting the GeV source as  an extended PWN does not necessarily require the TeV source be dominated the PWN. The TeV source can be interpreted as  other sources, such as an SNR.

In the PWN  scenario, the GeV emission is produced by diffusing electrons injected from the pulsar via IC scattering on the interstellar radiation. To study this possibility, we calculate the spatial template under the isotropic diffusion model. The details of the calculation on the gamma-ray emission and their spatial distribution can be found in Xi et al. (2019) and Liu et al. (2019). The magnetic field strength in the surrounding ISM is assumed to be $B=5\mu$G, and the interstellar radiation field (ISRF) consists of, in addition to CMB, a 20\,K far-infrared radiation with energy density of $0.4\,\rm eVcm^{-3}$, a 1300 K near-infrared radiation with energy density of $0.6\,\rm eVcm^{-3}$ and a 6500\,K optical radiation with energy density of $0.2\,\rm eVcm^{-3}$ following the ISRF model by Popescu et al. (2017). To model the injection history of electron/positron pairs from the pulsar, we assume an initial rotation period of $15\,$ms and a braking index of 3 for the pulsar, which then results in a pulsar age of $t_{\rm age}\simeq 140\,$kyr. A fraction of $\eta_e$ of the spin-down luminosity of the pulsar is assumed to be converted into the energy of pairs which are injected in a power-law spectrum with a slope $p$. We then estimate the required diffusion coefficient to be $D\simeq  10^{27}(E/1\rm TeV)^{1/3}cm^{-2}s^{-1}$ in order to explain the spatial extension of the GeV emission within the pulsar's age, given the best-fit $\sigma=0.42^\circ$ in the 2D Gaussian template which corresponds to 30\,pc at a nominal distance of 4.5\,kpc. This diffusion coefficient is about two orders of magnitude lower than the ISM value, consistent with the value found for the Geminga TeV halo (Abeysekara et al. 2017b). Since the position of the maximum TS is $\sim0.15\degr$ from the pulsar in the excess map, we have taken into account of the proper motion of the pulsar by assuming a plane-of-sky velocity $=0.15^\circ/t_{\rm age}\simeq 3.85\rm mas/yr$ or $83\rm km/s$.

Depending on whether the measured TeV emission has the same origin as the GeV emission, we obtain two templates with different  spectral indices (denoted by $p$) for the injected electron/positron pairs ($dN_e/d\gamma_e\propto \gamma_e^{-p}$), and the predicted flux is compared with data in Figures ~\ref{fig:1912-pwn} and ~\ref{fig:1912-pwn+snr}. {We generate the diffusion templates using the mapcube file, which is a 3 dimensional FITS map allowing arbitrary spectral variation as a function of sky position and cut at 2$\degr$.}
For $p=1.6$ and $\eta_e=0.06$, the measured TeV flux can be fitted simultaneously, with the TS value being 64.3 ($8.0\sigma$) for the GeV source. In the other template, a softer injection spectrum with $p=2.0$ is assumed, and the TS value is 67.5 ($8.2\sigma$) for the GeV source.
As the significance ($\approx\sqrt{TS}$) of the signal for these two diffusion templates is comparable to that of the Gaussian and disk templates, a possible physical explanation is that the GeV source is an extended PWN powered by PSR~J1913+1011.

\section{Discussions on the origin of the GeV/TeV source}
In this section, we discuss the possible nature of the GeV and TeV emission. One scenario for the GeV and TeV emissions is that they both arise from a PWN,  produced
by diffusing electrons that up-scatter the CMB and ambient photons.  The advantage of this scenario is that it can naturally explain the fact that the size of the GeV emission is significantly larger than that of the TeV emission measured by HESS. As shown by Fig. 5, the angular profile of TeV emission in the PWN scenario is steeper than that of the GeV emission due to stronger radiative cooling for TeV-emitting electrons, i.e., the intensity of TeV emission decreases more quickly with the distance to the center, which may explain the more compact size of the TeV emission. Indeed, some known PWNe,  such as HESS J1825-137,  show a clear energy-dependent morphology, with smaller sizes at higher energies (H.E.S.S collaboration 2019; Liu \& Yan 2019).   Modeling of the spectral energy distribution (SED) with this scenario is shown in Fig.~\ref{fig:1912-pwn}. The hard electron index ($p=1.6$) is not unusual for PWNe.  However, this scenario  can not explain the possible shell-like  morphology of the TeV emission, as suggested by recent  HESS observations (H.E.S.S. Collaboration et al. 2018), since the intensity in the PWN scenario is expected to decrease smoothly with the distance to the center. Since the field of view (FOV) of HESS is relatively small, a more clear angular profile of the TeV emission would require  observations by larger-FOV telescopes, such as HAWC and LHAASO.

{If the shell-like morphology is true, the TeV source may originate from an SNR (H.E.S.S collaboration 2018; Su et al. 2017). Given the relatively short radiative energy loss time of electrons producing TeV emission, it has been suggested that the shell-like TeV emission may
be due to hadronic interactions between high-energy protons accelerated by the SNR and the surrounding high-density molecular gas (Su et al. 2017). The GeV source can still be interpreted as an extended PWN in this scenario, as long as the TeV counterpart of the extended GeV PWN is dim, being outshone by the TeV SNR emission. The SED modeling with these two components (i.e., PWN plus SNR) is shown in Fig. 4. With a total energy of protons above 1 GeV of $3\times 10^{49}(n/5{\rm cm}^{-3})^{-1}$ erg, where $n$ is the mean ambient density, the hadronic emission from the SNR contributes significantly to the observed TeV emission. The maximum energy of the SNR-accelerated protons needs to be $E_M > $100TeV in this scenario. Such a high maximum energy is challenging for an old SNR with an age of 170 kyr. One possible solution is that these energetic protons were  accelerated in the early stage of the SNR evolution and they are still trapped by molecular clouds, giving rise to the observed TeV emission (Zeng et al. 2019).}


\section{2HWC J1831-098 and 2HWC J2031+415}

Two other TeV halo candidates (2HWC J1831-098 and 2HWC J2031+415) have been reported by Linden et al. (2017).
2HWC J1831-089 is an extended source (0.9$\degr$ disk) reported to be associated with the TeV source HESS J1831-089, a candidate PWN powered by the pulsar PSR J1831-0952 (Linden et al. 2017).  2HWC J2031+415 (0.7$\degr$ disk) is associated with the TeV source J2031+4130  detected by HEGRA in 2002, a PWN first reported as an unidentified source in TeV band (Aharonian et al. 2002). We apply the same \textit{Fermi-LAT} analysis to these sources. We found no  significant excess in GeV emission by testing different spatial templates  for 2HWC J1831-098 in the energy band 10-500 GeV, as shown in Fig. 5. An upper limit flux of $3.09\times10^{-12}$ erg cm$^{-2}$ s$^{-1}$ ($95\%$ confidence level) is obtained, assuming a spectral index of 2.0 and an uniform disk template of radius 0.9$\degr$ (centered at 2HWC J1831-098).
We compare this GeV flux limit measured by Fermi-LAT with the TeV flux of 2HWC J1831-098 and find that the upper limit is significantly lower than the TeV flux ($7.60\times10^{-12}$ erg cm$^{-2}$ s$^{-1}$). This may imply that a hard electron index is needed in the PWN scenario for the GeV and TeV emissions, which is quite similar to the Geminga TeV halo (Xi et al. 2019).

The source 2HWC J2031+415  is located within the region of the GeV-bright Cygnus Cocoon (distance of 0.664$\degr$). The diffuse emission in this region was fitted by a 2D Gaussian model with $\sigma=3.0\degr$ by the Fermi collaboration (The Fermi-LAT collaboration 2019).
As seen in the right panel of Fig. 5, after  subtracting the emission of the Cygnus Cocoon, no significant excess is found in the residual map.  Therefore, it is hard to identify  possible GeV emission from 2HWC J2031+415 in the region of the bright  Cygnus Cocoon. For this reason, we do not analyze this source further.

\section{Conclusions}

Using $\sim$10.4 years of \textit{Fermi-LAT} data, we discovered a new extended GeV source in the vicinity of the TeV source 2HWC J1912+099, which is also detected by HESS (HESS J1912+101). The best-fit extension of the GeV $\gamma$-ray emission is $0.85\degr\pm0.08\degr$ using the uniform-disk template,
significantly larger than the size of the TeV emission measured by HESS ($\sim$0.5$\degr$). Considering also that there is a middle-aged radio pulsar PSR~J1913+1011 at the center of the GeV/TeV emission, we studied the possibility that the GeV/TeV source is an extended PWN, i.e., a TeV halo. Then we test various diffusion templates, which assumes that the GeV emission is  produced by diffusing electrons, and find that the diffusion spatial templates give an almost equally good fit to the data as the uniform-disk spatial template and Gaussian spatial template. Thus, the  GeV source can be interpreted as an extended PWN. However,  the possible shell-like morphology of the TeV emission, if true, is inconsistent with  this scenario.

{Alternatively, the shell-like TeV morphology may be attributed to emission from an SNR. In this scenario, the TeV emission is likely produced by ions interacting with the surrounding high density molecular gas (Su et al. 2017), given the very short radiative loss time for  TeV-emitting electrons. By studying a sample of 34 $\gamma$-ray SNRs, Zeng et al. (2019) found that the acceleration of TeV particles usually stops before the end of the Sedov phase when the remnant is  $\la 10$ kyr, which is much shorter than the age (170kyr) of the system. One possible solution to this dilemma is that these energetic protons were  accelerated in the early stage of the SNR evolution, and they are still trapped by molecular clouds, producing the observed TeV emission. The GeV source can still be interpreted as an extended PWN with its TeV emission outshone by the SNR. This is possible if the electron spectrum of the PWN is soft. A precise measurement of the angular profile of the TeV emission in the future will be useful to diagnose the electron diffusion in the TeV halo scenario. Meanwhile, future deeper multi-wavelength observations to search for the low-energy
counterparts of a potential SNR will be useful to test the SNR scenario.}


\section*{Acknowledgments}
We thank the referee for the constructive report and Hui Zhu for useful discussions. The work is supported by the National Key R \& D program of China under the grant 2018YFA0404203 and the NSFC grants 11625312 and 11851304.

\clearpage

\begin{figure*}
\includegraphics[angle=0,scale=0.32]{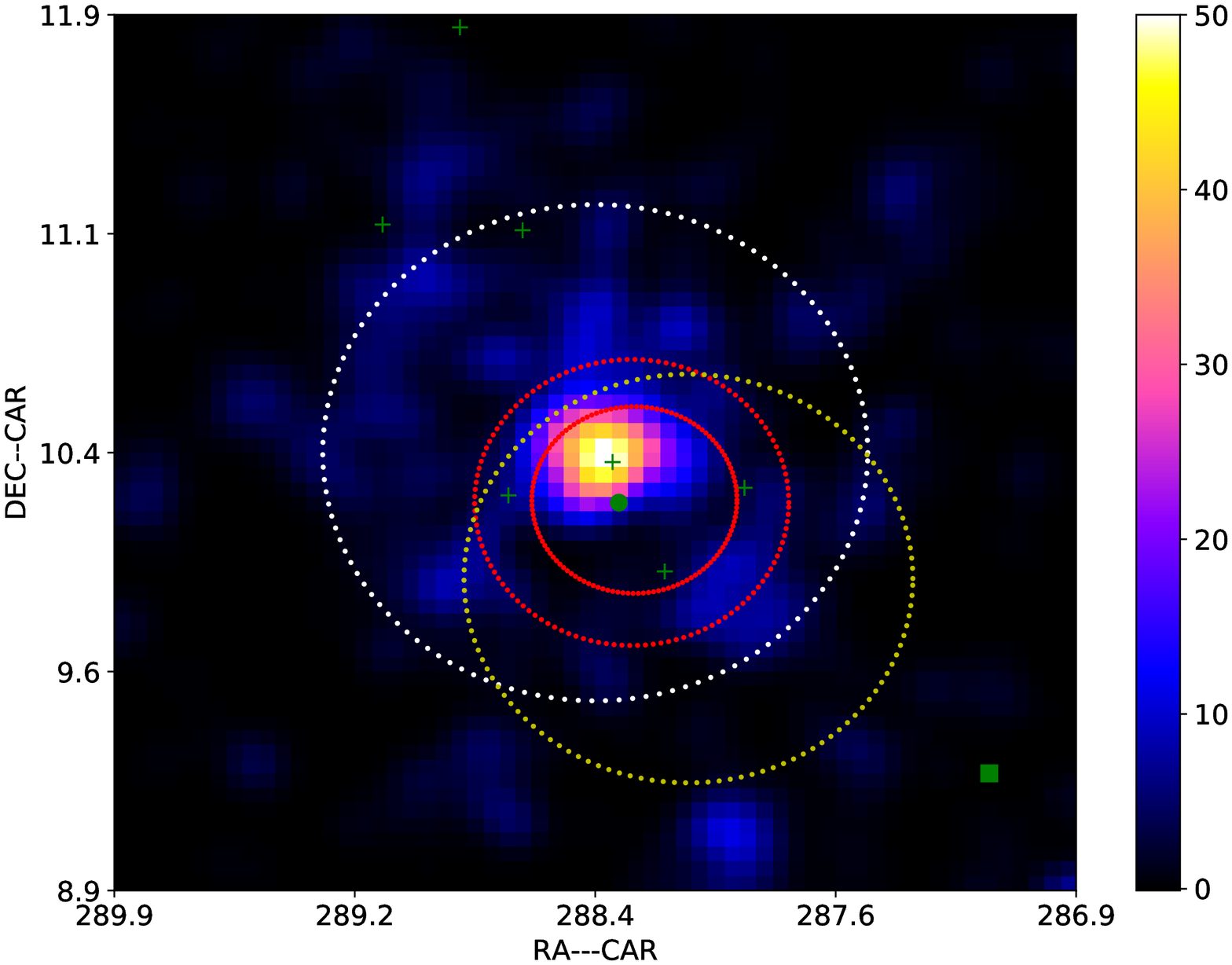}
\includegraphics[angle=0,scale=0.32]{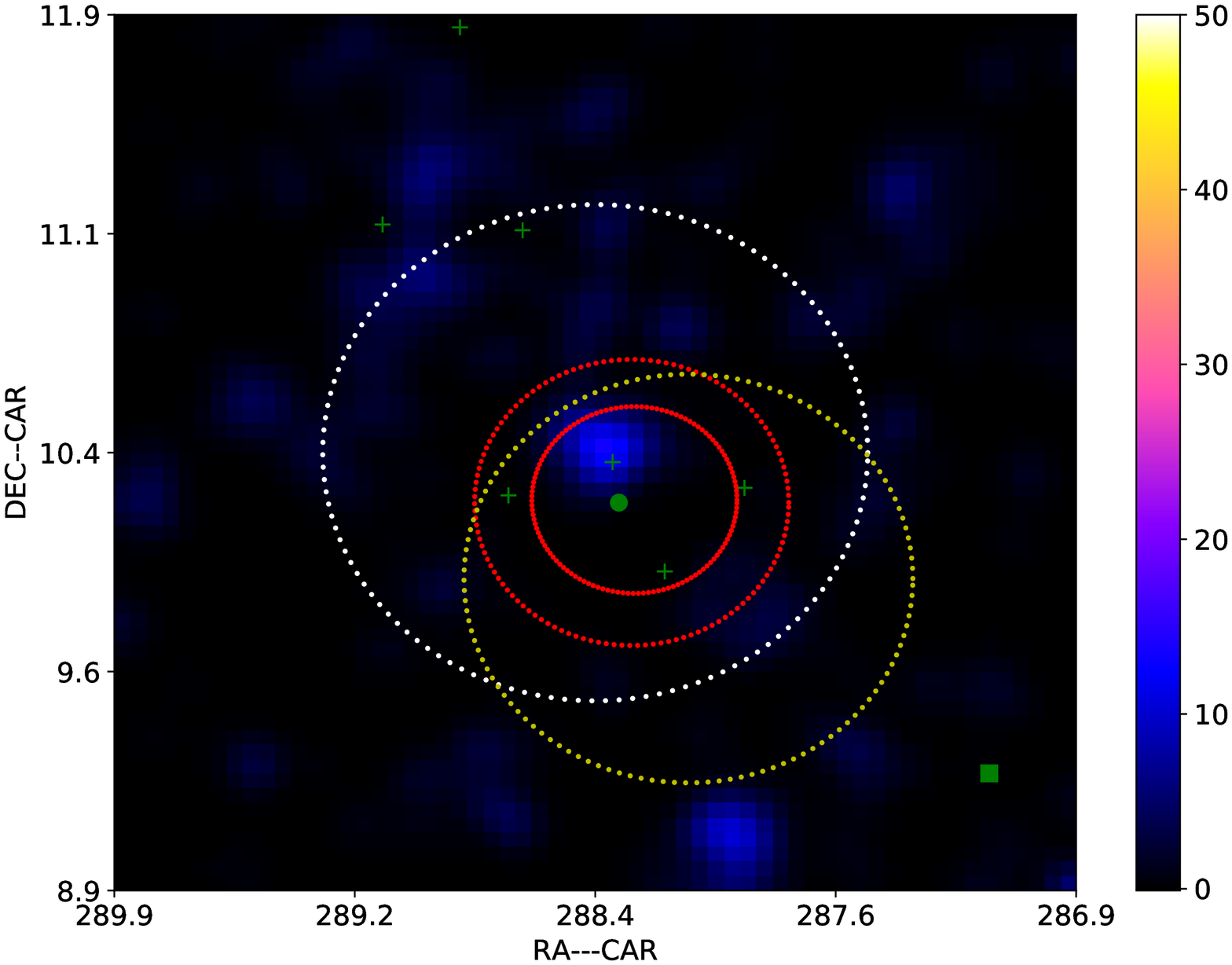}\\
\caption{$3\degr \times 3\degr$ TS maps of the HESS J1912+101 region  in the energy band 10-500 GeV, with pixel size corresponding to $0.05\degr \times 0.05\degr$. The green crosses represent the positions of the 4FGL point sources and the green square represents the extended source G42.8+0.6. The two red dashed circles indicate the outer and inner radii of the  shell model for the TeV source HESS J1912+101(H.E.S.S collaboration 2018), while the yellow and the white dashed circles indicate the extents of 2HWC J1912+099 (Abeysekara et al. 2017) and the GeV emission (uniform-disk template) in this work, respectively. The position of the PSR J1913+1011 is shown with a green dot.
Left panel: excess map with all the background components subtracted, including the diffuse Galactic background, isotropic background,  extended sources and point sources in 4FGL.
Right panel: the residual map with {an additional extended GeV source being subtracted, assuming a disk template for the GeV emission}.
All the maps have been created for a pixel size of 0.05, smoothed by gaussian kernel($\sigma=0.35\degr$).
The color bar represents the value of TS per pixel.}\label{}
\end{figure*}

\begin{figure*}
\includegraphics[angle=0,scale=0.3]{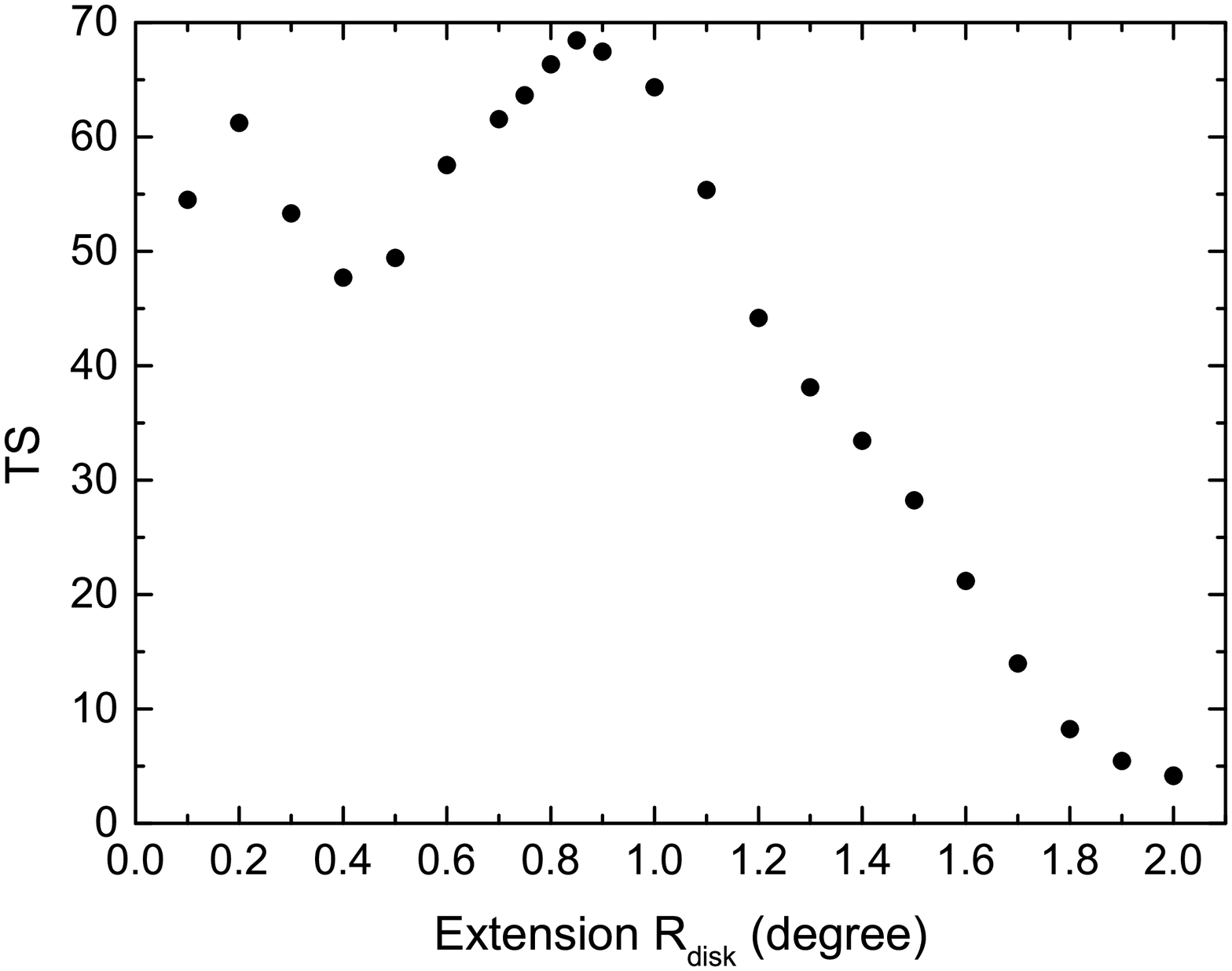}
\includegraphics[angle=0,scale=0.3]{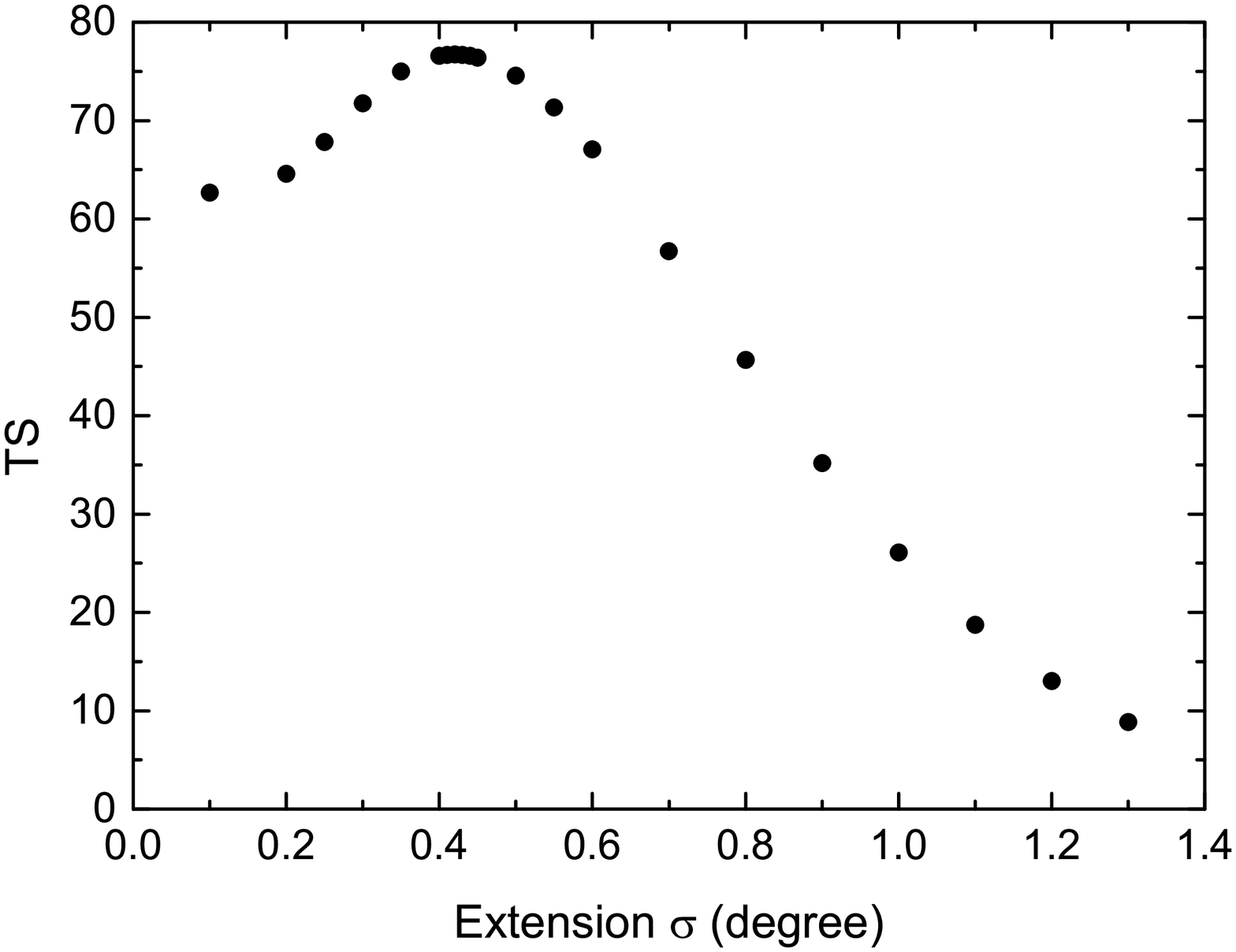}
\caption{The significance of the GeV source assuming  uniform-disk(left) and 2D-Gaussian(right) models with various $R_{disk}$ and $\sigma$, respectively.  The maximum TS of the uniform-disk model is at $R_{disk}=0.85\pm0.08\degr$ with TS=68.5. The maximum TS of the 2D Gaussian model is at $\sigma=0.42\pm0.03\degr$, corresponding to 68\% containment radius $R_{68}=0.63\pm0.05\degr$, with TS=76.7. }\label{}
\end{figure*}

\begin{figure*}
\includegraphics[angle=0,scale=0.5]{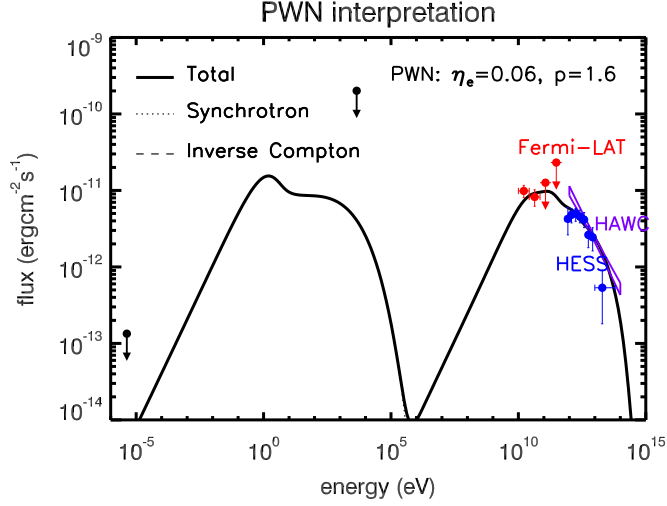}
\caption{Modeling the multi-band spectrum of the source with the electron synchrotron plus inverse-Compton emission in the  PWN scenario. The data points and upper limits (3$\sigma$ confidence level) of the GeV emission (10 GeV - 500 GeV), obtained using the diffusion templates, are shown as red points. The blue data points represent the H.E.S.S energy flux spectra of HESS J1912+101. The purple butterfly shows the best fit power-law model with $\Gamma$=2.64$\pm$0.06 of 2HWC J1912+099. The black data points represent the upper limits of radio emission (Reich $\&$ Sun 2019) and X-ray emission (Chang et al. 2008). Model parameters are $P_0=15\,$ms, $n=3$, $D=10^{27}(E/1\,{\rm TeV})^{1/3}\rm cm^2s^{-1}$, $\eta_e=0.06$, and $p=1.6$. }\label{fig:1912-pwn}
\end{figure*}

\begin{figure*}
\includegraphics[angle=0,scale=0.5]{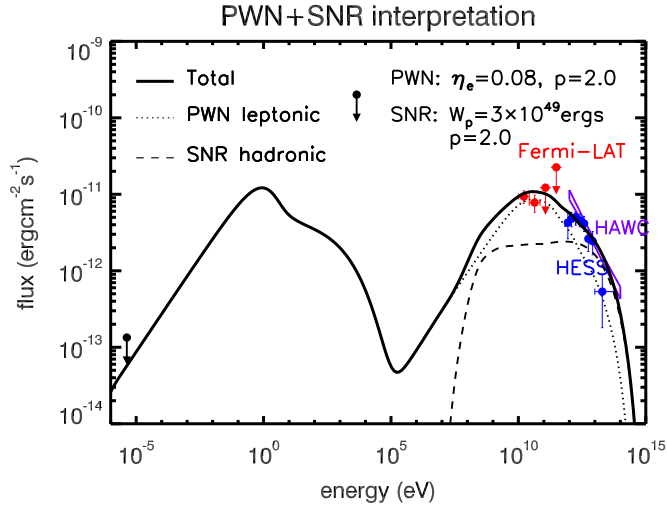}
\caption{Modeling the multi-band spectrum of the  source in the joint model, where the GeV emission is mainly produced by a PWN and a  SNR contributes significantly to the TeV emission. The data points are the same as that in Fig.3. The model parameter values for the PWN are the same as that in Fig. 3 except for $\eta_e=0.08$ and $p=2.0$.  Model parameters for the SNR are $E_p=3\times10^{49}{\rm erg}$, $p=2$ and a gas density of $n=5 {\rm cm^{-3}}$. }\label{fig:1912-pwn+snr}
\end{figure*}

\begin{figure*}
\includegraphics[angle=0,scale=0.5]{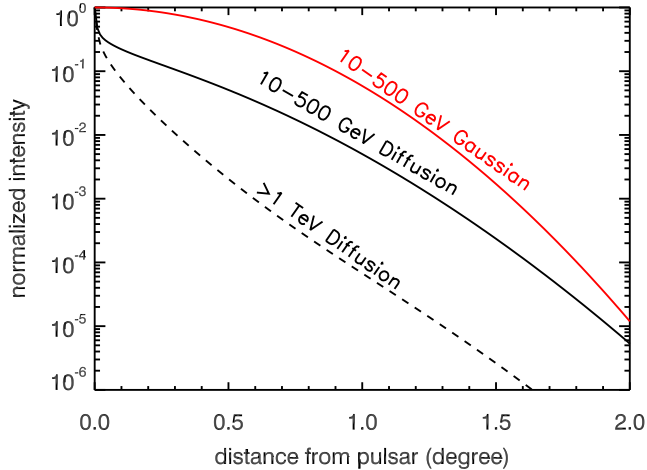}
\caption{The surface brightness profiles in $10-500\,$GeV and $>1\,$TeV for the diffusion template and for the 2D Gaussian template $10-500\,$GeV. }\label{}
\end{figure*}

\begin{figure*}
\includegraphics[angle=0,scale=0.22]{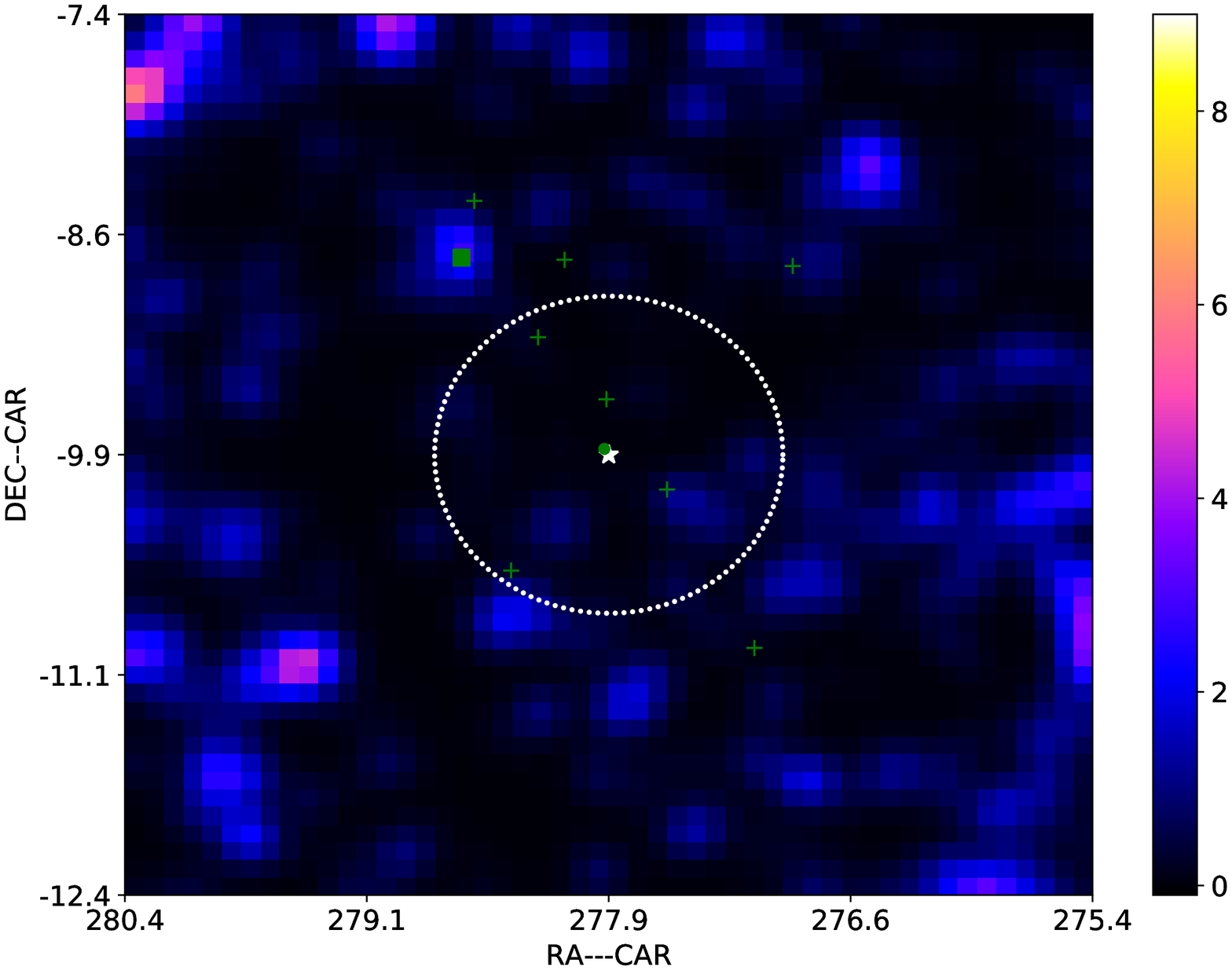}
\includegraphics[angle=0,scale=0.22]{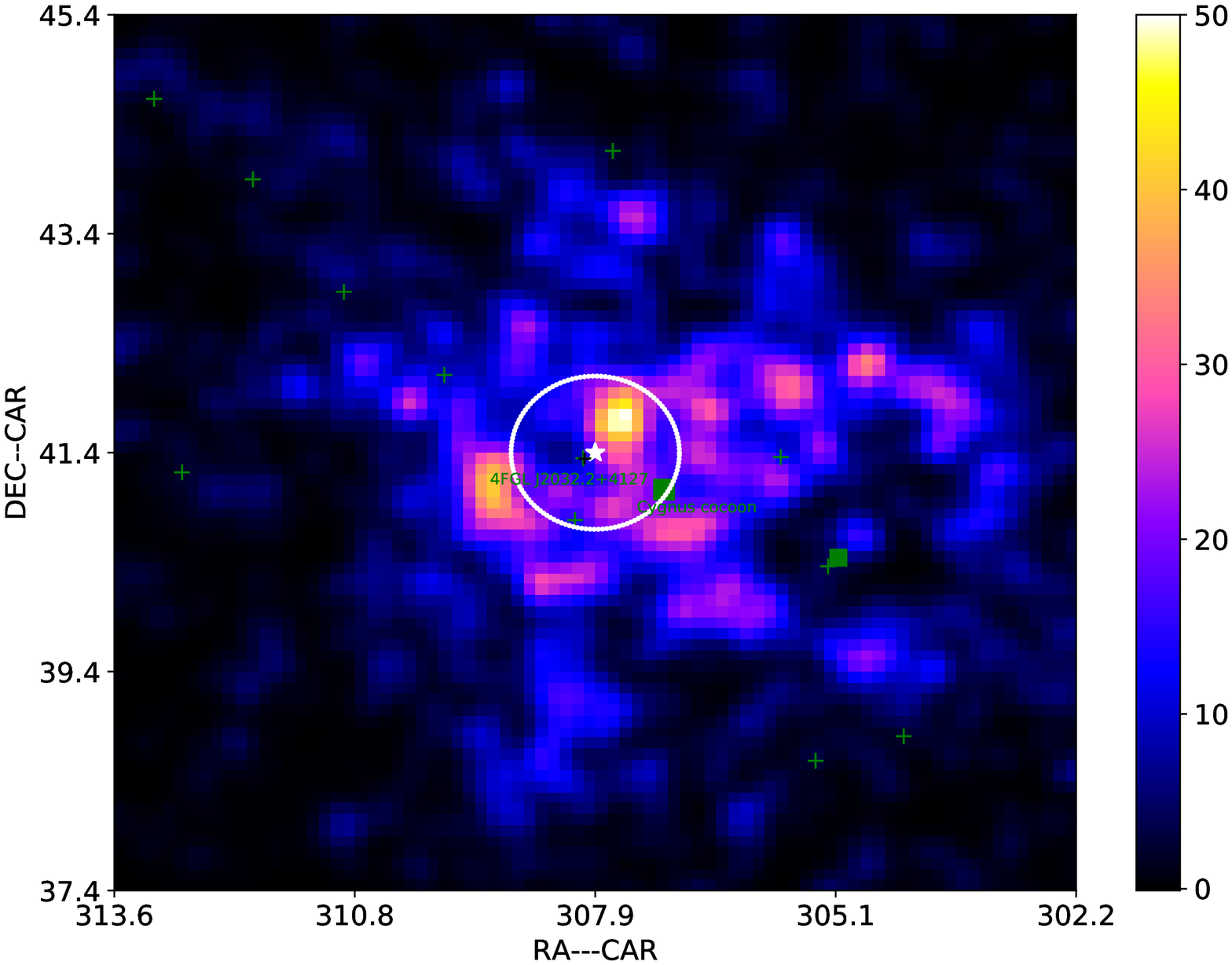}
\includegraphics[angle=0,scale=0.22]{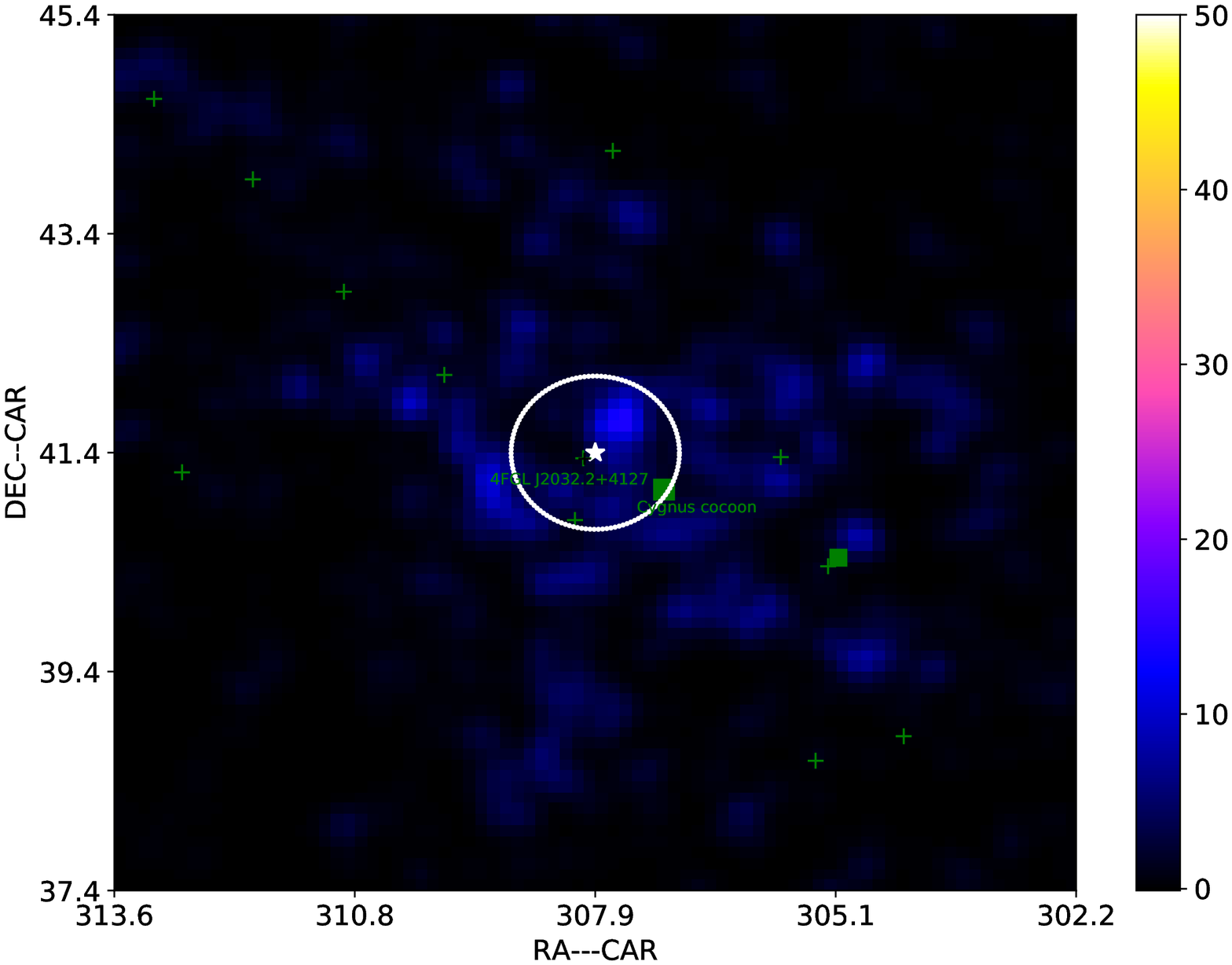}\\
\caption{Left panel: the excess map around the region of 2HWC J1831-089 with all the background components being subtracted. No significant excess emission is found. The white star and dashed circle represent center and edge of the TeV source 2HWC J1831-089. The pulsar PSR J1831-0952 is shown by a green dot. Middle panel: the excess map around the region of 2HWC J2031+415 with all the background components subtracted except the cygnus cocoon (the center of the cygnus cocoon is shown by a green square). Right panel: the excess map with all the background components, including the cygnus cocoon, subtracted. The white star and dashed circle represent the center and edge of the TeV source 2HWC J2031+415. The position of the $\gamma$-ray pulsar PSR J2032+4127 is shown by a black cross. Note that the source 4FGL J2032.2+4127 is associated with PSR J2032+4127 (The Fermi-LAT collaboration(2019)). All the maps have been created for a pixel size of 0.1, smoothed by gaussian kernel($\sigma=0.7\degr$).
The color bar represents the value of TS per pixel.}\label{}
\end{figure*}

\begin{table*}
\caption{TS values and spectral information of the GeV source for  different spatial templates}
\begin{tabular}{lccccccc}
\hline
\hline
Spatial Template & Energy Flux&  Spectral Index  &TS \\
 &($\times 10^{-11}{\rm \ erg\ cm^{-2}\ s^{-1}}$) & & &\\
 \hline
 PS\footnotemark[1] & 0.37 $\pm$0.10 & 3.05 $\pm$0.20 & 45.3\\
\\
HESS Map\footnotemark[2] & 1.62 $\pm$ 0.39 & 1.96 $\pm$ 0.10 & 43.2  \\
\\
Uniform disk ($0.85\degr$) & 2.53 $\pm$0.47 & 2.41 $\pm$0.18 &68.5 \\
\\
Gaussian ($\sigma=0.42\degr$) & 2.69 $\pm$ 0.44 & 2.44 $\pm$ 0.10 & 76.7 \\
\\
Diffusion 1\footnotemark[3]  & 3.18 $\pm$ 0.45 & - & 64.3 \\
\\
Diffusion 2\footnotemark[4]  & 2.71 $\pm$ 0.37 & - & 67.5 \\
\\
\hline
\hline
\end{tabular}
\par
 Notes:\\
\footnotemark[1]{PS corresponds to the point source model with a power-law spectrum.}\\
\footnotemark[2] {HESS Map represents the TeV shell model with outer radius of 0.49$\degr$ and inner radius of 0.32$\degr$} (H.E.S.S collaboration 2018).\\
\footnotemark[3]{The parameters used for the diffusion template 1 are $P_0=15\,$ms, $n=3$, $D=10^{27}(E/1\,{\rm TeV})^{1/3}\rm cm^2s^{-1}$, $\eta_e=0.06$, $p=1.6$. See also Section~3 for details.}\\
\footnotemark[4]{The parameters used for the diffusion template 2 are the same as those in the diffusion template 1, except $\eta_e=0.09$ and $p=2.0$.}\\
\label{tab1}
\end{table*}


\begin{thebibliography}{99}


\bibitem[Abdo et al.(2009)]{2009ApJ...700L.127A} Abdo, A.~A., Allen, B.~T., Aune, T., et al.\ 2009, \apjl, 700, L127


\bibitem[Abeysekara et al.(2017a)]{2017ApJ...843...40A} Abeysekara, A.~U., Albert, A., Alfaro, R., et al.\ 2017a, \apj, 843, 40

\bibitem[Abeysekara et al.(2017b)]{}Abeysekara, A.~U., Albert, A., Alfaro, R., et al. 2017b, Science, 358, 911


\bibitem[Ackermann et al.(2017)]{2017ApJ...843..139A} Ackermann, M., Ajello, M., Baldini, L., et al.\ 2017, \apj, 843, 139


\bibitem[Aharonian et al.(2002)]{2002A&A...393L..37A} Aharonian, F., Akhperjanian, A., Beilicke, M., et al.\ 2002, \aap, 393, L37


\bibitem[Aharonian et al.(2008)]{2008A&A...484..435A} Aharonian, F., Akhperjanian, A.~G., Barres de Almeida, U., et al.\ 2008, \aap, 484, 435


\bibitem[Atwood et al.(2009)]{2009ApJ...697.1071A} Atwood, W.~B., Abdo, A.~A., Ackermann, M., et al.\ 2009, \apj, 697, 1071


\bibitem[Chang et al.(2008)]{2008ApJ...682.1177C} Chang, C., Konopelko, A., \& Cui, W.\ 2008, \apj, 682, 1177

\bibitem[Condon et al. (1998)]{} Condon, J. J., Cotton, W. D., Greisen, E. W., et al. 1998, AJ, 115, 1693


\bibitem[Di Mauro et al.(2019)]{2019arXiv190305647D} Di Mauro, M., Manconi, S., \& Donato, F.\ 2019, arXiv:1903.05647

\bibitem[Helfand et al. (2006)]{}Helfand, D. J., Becker, R. H., White, R. L., Fallon, A., \& Tuttle, S. 2006, AJ, 131, 2525


\bibitem[H.E.S.S.~Collaboration et al.(2018)]{2018A&A...612A...8H} H.E.S.S.~Collaboration, Abdalla, H., Abramowski, A., et al.\ 2018, \aap, 612, A8

\bibitem[H.E.S.S.~Collaboration et al.(2018)]{} H.E.S.S.~Collaboration, Abdalla, H., Abramowski, A., et al. \aap, 2019, 621, A116
\bibitem[Lande et al.(2012)]{2012ApJ...756....5L} Lande, J., Ackermann, M., Allafort, A., et al.\ 2012, \apj, 756, 5


\bibitem[Linden et al.(2017)]{2017PhRvD..96j3016L} Linden, T., Auchettl, K., Bramante, J., et al.\ 2017, \prd, 96, 103016


\bibitem[Liu et al.(2019)]{2019ApJ...875..149L} Liu, R.-Y., Ge, C., Sun, X.-N., et al.\ 2019, \apj, 875, 149


\bibitem[Morris et al.(2002)]{2002MNRAS.335..275M} Morris, D.~J., Hobbs, G., Lyne, A.~G., et al.\ 2002, \mnras, 335, 275


\bibitem[Popescu et al. (2017)]{}
Popescu, C. C., Yang, R., Tuffs, R. J., et al., 2017, MNRAS, 470, 2539


\bibitem[Reich \& Sun(2019)]{2019RAA....19...45R} Reich, W., \& Sun, X.-H.\ 2019, Research in Astronomy and Astrophysics, 19, 045


\bibitem[The Fermi-LAT collaboration(2019)]{2019arXiv190510771T} The Fermi-LAT collaboration 2019, arXiv:1905.10771

\bibitem[Su et al. (2017)]{} Su, Y., Zhou, X., Yang, J., et al., 2017, ApJ, 845, 48


\bibitem[Xi et al.(2019)]{2019ApJ...878..104X} Xi, S.-Q., Liu, R.-Y., Huang, Z.-Q., Fang, K., \& Wang, X.-Y.\ 2019, \apj, 878, 104

\bibitem[Zeng et al. (2019)]{}Zeng, H. D. Xin, Y. L., Liu, S. M., 2019, ApJ, 874, 50
\end{thebibliography}
\end{document}